%% file: main.tex
\documentclass[conference,9pt]{IEEEtran}
\usepackage[usenames,dvips]{color}
\usepackage{multirow}
\usepackage{booktabs}
\usepackage{amsmath}
\usepackage{amsfonts}
\usepackage{multirow}
\usepackage{epsfig}
\usepackage{psfrag}
\usepackage{subfigure}
\usepackage{pifont}
\usepackage{cite}
\usepackage{url}
\newcounter{numberlistc}
\usepackage{algorithm}
\usepackage{algorithmic}
\usepackage{graphicx}
\graphicspath{ {./figs/} }

\usepackage[usenames]{color}
\usepackage{amssymb,amsmath}
\usepackage{amsthm}
\usepackage{subfigure}

\usepackage{amssymb} 
 \usepackage{amsmath}

\newcounter{itemlistc}

\usepackage{setspace}

\usepackage[font=small, labelfont=bf]{caption}
\begin{document}
\date{}

\title{Run-Time Accuracy Reconfigurable Stochastic Computing for
  Dynamic Reliability and Power Management}

\author{
\IEEEauthorblockN{Shuyuan Yu\IEEEauthorrefmark{1}, Han
  Zhou\IEEEauthorrefmark{1}, Shaoyi Peng\IEEEauthorrefmark{1},
  Hussam Amrouch\IEEEauthorrefmark{2}, Joerg Henkel\IEEEauthorrefmark{2},
   Sheldon X.-D. Tan\IEEEauthorrefmark{1}}
\IEEEauthorblockA{\IEEEauthorrefmark{1}
Department of Electrical and Computer Engineering, 
University of California, Riverside, CA 92521,
stan@ece.ucr.edu}
\IEEEauthorblockA{\IEEEauthorrefmark{2}
Karlsruhe Institute of Technology, Chair for Embedded Systems (CES), Karlsruhe, Germany}
%\IEEEauthorblockA{\IEEEauthorrefmark{3}
% Mentor Graphics Corporation,
%   Fremont, CA 94538}
%   \thanks{This work is supported in part by NSF grant under
%     No. CCF-1017090, in part by NSF Grant under No. CCF-1255899, in
%     part by Semiconductor Research Corporation (SRC) grant under
%     No. 2013-TJ-2417.}
}

\maketitle
 
\begin{abstract}
  In this paper, we propose a novel accuracy-reconfigurable stochastic
  computing (ARSC) framework for dynamic reliability and power
  management. Different than the existing stochastic computing works,
  where the accuracy versus power/energy trade-off is carried out in
  the design time, the new ARSC design can change accuracy or
  bit-width of the data in the run-time so that it can accommodate the
  long-term aging effects by slowing the system clock frequency at
  the cost of accuracy while maintaining the throughput of the
  computing. We validate the ARSC concept on a discrete cosine
transformation (DCT) and inverse DCT designs for image
compressing/decompressing applications, which are implemented on
Xilinx Spartan-6 family XC6SLX45 platform. 
  % Further more, the
  % proposed reconfigurable SC method can provide new knob to
  % dynamically regulate the active power of a chip as one can scale the
  % frequency in much arange (compared to traditional voltage and
  % frequency scaling techniques) to trade the accuracy for power in a
  % progressive way.
Experimental results shows that the new design can easily mitigate the
long-term aging induced effects by accuracy trade-off while
maintaining the throughput of the whole computing process using simple
frequency scaling. We further show that one-bit precision loss for
input data, which translated to 3.44dB of the accuracy loss in term
of Peak Signal to Noise Ratio for images, we can sufficiently
compensate the NBTI induced aging effects in 10 years while maintaining
the pre-aging computing throughput of 7.19 frames per second. At the same
time, we can save 74\% power consumption by 10.67dB of accuracy loss. 
The proposed ARSC computing framework also allows much aggressive frequency scaling, which can lead to
order of magnitude power savings compared to the traditional dynamic
voltage and frequency scaling (DVFS) techniques.

\end{abstract}

\input intro.tex

\input sc_review.tex

%\input motivation.tex

\input architecture.tex

%\input strategy.tex

\input hard.tex

\input results.tex

\section{Conclusion}
\label{sec:conclusion}

In this paper, we have proposed a novel accuracy-reconfigurable
stochastic computing (ARSC) framework for dynamic reliability and
power management. The new ARSC design can dynamically change accuracy
via bit-width change of the data. In this way, the new method can
accommodate the long-term aging effects by slowing the system clock
frequency at the cost of accuracy while maintaining the throughput
of the computing. We designed and validated the ARSC-based discrete
cosine transformation (DCT) and inverse DCT designs for image
compressing/decompressing applications on the Xilinx Spartan-6 family
XC6SLX45 platform.  Experimental results show that one can easily
mitigate the long-term aging effects by accuracy reduction while
maintaining the throughput of the whole computing process using simple
frequency scaling. In our example, we show that one-bit precision loss
for the input data, which translated to 3.44dB of the accuracy loss in
term of Peak Signal to Noise Ratio (PSNR) for images, one can
sufficiently compensate the NBTI induced aging effects in 10 years
while maintaining the pre-aging computing throughput of 7.19 frames
per second. At the same time, one can save 74\% power consumption by 
10.67dB of accuracy loss. The
proposed ARSC computing framework allows much aggressive frequency
scaling, which can lead to order of magnitude power savings compared
to the traditional dynamic voltage and frequency scaling (DVFS)
techniques.

\begin{scriptsize}
  \bibliographystyle{ieeetr}
\bibliography{../../bib/emergingtech,../../bib/thermal_power,../../bib/mscad_pub,../../bib/interconnect,../../bib/stochastic,../../bib/simulation,../../bib/modeling,../../bib/reduction,../../bib/misc,../../bib/architecture,../../bib/reliability,../../bib/reliability_papers,../../bib/neural_network,../../bib/approximate_comp}
\end{scriptsize}

\end{document}

%% file: intro.tex
\section{Introduction}
\label{sec:intro}

One of the important paradigm changes for today's emerging computing
workloads such as deep learning, AI, computer vision, imaging and audio
processing is that accurate computing becomes less important as those
applications are much more error tolerant with analog-like outputs for
human interaction.  As a result, accuracy can be traded off to improve
hardware footprint, power/energy efficiencies via so-called
approximation computing.  One important approach for approximate
computing is by means of stochastic computing (SC), in which the value
is presented as the signal probability in a bit stream instead of
traditional binary number~\cite{AlaghiQian:TCAD'18}. SC is shown to
have better error resilience, progressive trade-off among performance,
accuracy and energy, as well as cheap implementation of complex
arithmetic operations.

\begin{figure}[htp]
\centering
\includegraphics[width=0.8\columnwidth]{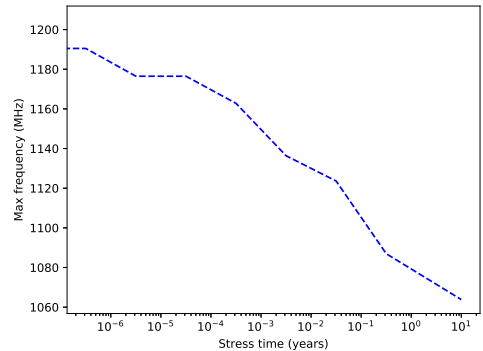}
\caption{The maximum working frequency decreases over years because of
aging.}
\label{fig:aging}
\end{figure}

On the other hand, today's digital systems are built on less reliable
devices and less robust interconnects as technology node advances.
The major reliability effects for VLSI chips include bias temperature
Instability (BTI), hot carrier injection (HCI) for CMOS devices,
electromigration (EM) and time dependent dielectric breakdown
(TDDB) for interconnects and dielectrics, which are the major
consideration for the aging
effects~\cite{Faiil_device'02,ITRS_reliab'03}. Fig.~\ref{fig:aging}
shows how BTI affects maximum frequency of a discrete cosine
transformation (DCT) design based on the Nangate 45nm
degradation-aware standard cell library from Karlsruhe Institute of
Technology (KIT)~\cite{KIT_LIB}.  Those aging and long-term
reliability effects such are getting worse with shrinking feature
sizes and future chips will show signs of aging much faster than the
previous generations~\cite{TanAmrouch:2017int}.  To mitigate the
increasing reliability and resiliency problems, traditional long-term
reliability and aging analysis mainly focus on the reliability
optimization at the design time of the system and physical
level~\cite{TanTahoori:Book'19}.  Recently using less accurate
computing to compensate the NBTI-induced long-term aging effects have
been proposed~\cite{Amrouch:DAC'17}. However this method is targeted
at the design time so that sufficient margins can be allocated in
advance.

% We need to show a few figures for the BTI and EM induced performance
% and frequency degrations over time.

% We also need to discuss existing reconfigurable SC is mainly for
% function reconfiguration, not for accuracy.

%We also need to review a bit of traditional SC, the last BI-SC idea,
% which significant reduce SC latency.
On the other hand, stochastic computing has been emerging a new
computing paradigm due to its low-cost and error-resilient features.
% Fig.~\ref{fig:conv_sc_mul} shows the conventional stochastic (SC)
% computing multiplier, where the number or stochastic number (SN), is
% represented by a bit-stream, whose signal probability, or frequency
% of bit `1', determines its value. Naturally, the value is defined in
% the range $[0,1]$, called unipolar, or over $[-1,1]$ called
% bipolar. For instance, in Fig.~\ref{fig:conv_sc_mul}, the number $X$
% represents $4/8$ as we have four bit `1' in the 8 bit bit-stream.
One of the
major benefits for SC is that many arithmetic operations such as
multiplication can be simply implemented by {\it AND} operation (or {\it XNOR}
%gate for bipolar) as shown in this figure.
gate for bipolar).
SC has been applied to error-correcting codes~\cite{NaderiMannor:IEEE'11}, 
image processing~\cite{LiLilja:IEEE'13}, and recently deep neural networks
(DNNs)~\cite{KimKim:DAC'16, SimNguyen:ASPDAC'17, SimLee:DAC'17,
  HojabrGivaki:DAC'19}.
%\begin{figure}[htp]
%\centering
%\includegraphics[width=0.9\columnwidth]{figs/conv_sc_mul.eps}
%\caption{Conventional SC number and multiplier.}
%\label{fig:conv_sc_mul}
%\vspace{-0.2in}
%\end{figure}

Traditional SC, however, suffers long computing time and high
randomness of the stochastic numbers for accuracy. As a result, many
research works have been proposed to mitigate those shortcomings such as
high-quality random number generators (RNGs) that exhibit zero or
close to zero correlation, including low-discrepancy
sequences~\cite{LiuHan:DATE'17}, bit scrambling
methods~\cite{Neugebauer:DSD'17,KimLee:ASPDAC'16}.  Recently, a more
efficient and also accurate SC multiplier was proposed to partially
mitigate the two mentioned problems in the traditional
SC~\cite{SimLee:DAC'17}. Instead of using an {\it AND} gate for the
multiplication of two bit-streams, the new multiplier essentially
counts the one in one bit-stream based on the value of another
bit-stream. Further more, the bit-stream to be counted can be
generated in a deterministic way. As a result, the whole design is
simplified into two counters and a simple bit-stream generator. In
this work, we call this design {\it counter-based SC multiplier}
(CBSC-Multiplier).  CBSC-Multiplier brings two important benefits:
first, it does not require the randomness of the two bit streams
anymore without loss of accuracy. Second, it can be faster than
traditional SC as it drops the requirement of counting all bits in a
bit-stream.

Based on those observations, in this paper, we propose a new
accuracy-reconfigurable stochastic computing (ARSC) technique for
dynamic long-term reliability management and more power efficient
computing. It leverage the latest CBSC computing frameworks for more
energy-efficient SC implementation.
Different than existing
stochastic computing works, where the accuracy versus power/energy
trade-off is carried out in the design time, the new stochastic
computing can change accuracy or bit-width of the data in the run-time
so that it can accommodate the long-term aging effects by slowing the
system clock frequencies at the cost of accuracy while maintaining the
throughput of the computing. As many emerging workloads are error
tolerant, the new accuracy-reconfigurable stochastic computing
essentially provides viable solution to mitigate the challenging
long-term reliability problems due to the increasing degradation
effects such as biased temperature instability (BTI) and
electromigration (EM) as technology advances. Further more, the
proposed reconfigurable SC method can provide new knob to dynamically
regulate the active power of a chip as one can scale the frequency in
much larger range (compared to traditional voltage and frequency
scaling techniques) to trade the accuracy for power in a progressive
way.

We validate the ARSC concept on a discrete cosine transformation (DCT)
and inverse DCT designs for image compressing/decompressing
applications, which are implemented on Xilinx Spartan-6 family
XC6SLX45 platform.  Experimental results shows that the new design can
easily mitigate the long-term aging induced effects by accuracy
trade-off while maintaining the throughput of the whole computing
process using simple frequency scaling. We further show that one-bit
precision loss for input data, which translated to 3.44dB of the
accuracy loss in term of Peak Signal to Noise Ratio for images, we can
sufficiently compensate the NBTI induced aging effects in 10 years
while maintaining the pre-aging computing throughput of 7.19 frames
per-second. At the same time, we can save 74\% power consumption by 
10.67dB of accuracy loss. The
proposed ARSC computing framework also allows much aggressive frequency
scaling, which can lead to order of magnitude power savings compared
to the traditional dynamic voltage and frequency scaling (DVFS)
techniques.

%% file: sc_review.tex
\section{Review of stochastic computing}

Stochastic computing (SC) provides an alternative way for arithmetic
computing when the exact results are not required. At the same time,
the SC based hardware can be designed with extremely low cost and low
power than traditional binary digital designs.

\subsection {Conventional stochastic computing:}
\begin{figure}[htp]
%\begin{wrapfigure}{L}{0.4\textwidth}
\centering
\includegraphics[width=0.8\columnwidth]{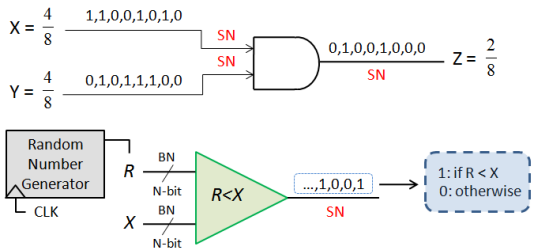}
\caption{Traditional SC number and multiplier.}
\label{fig:conv_sc_mul}
%\vspace{-0.1in}
% \end{wrapfigure}
\end{figure}

Fig.~\ref{fig:conv_sc_mul} shows the conventional stochastic computing
(SC)  multiplier, where the number or stochastic number, SN, is
represented by a bitstream, whose signal probability, or frequency of
'1', determines its value. Naturally, the value is defined in the
range $[0,1]$, called unipolar, or over $[-1,1]$ called bipolar. For
instance, Fig.~\ref{fig:conv_sc_mul}, the number $X$ represents $4/8$
as we have four '1' in the 8-bit bitstream. One of the major benefits for
SC is that the multiplication can be simply implemented by $AND$
operation as shown in this figure.

To generate the random number for SC, stochastic number generator,
SNG, which essentially converts binary number to stochastic number, takes
$n$-bit binary number and generates the random bitstream as shown
in the bottom part of Fig.~\ref{fig:conv_sc_mul}. SNG typically is
implemented by $n$-bit linear feedback shift register (LFSR) and $n$-bit comparator, which generates '1' if the random number is less than
the input binary number, and '0' otherwise.

%\begin{wrapfigure}{L}{0.4\textwidth}
%\begin{figure}[htp]
%\centering
%\includegraphics[width=0.8\columnwidth]{./conv_rng.eps}
%\caption{Traditional SC random number generator.}
%\label{fig:conv_sc_RNR}
%\vspace{-0.1in}
% \end{wrapfigure}
%\end{figure}

For stochastic computing unipolar encoding, the multiplication can be
done by $AND$ operation and for bipolar encoding, the multiplication is
achieved by $XNOR$
operation~\cite{Gaines:Book'69,AlaghiQian:TCAD'18}. The addition 
can be simply done by a multiplexer (MUX)~\cite{Gaines:Book'69,
  AlaghiQian:TCAD'18}. Finally, the resulting bit-stream can be
converted back to a binary number using a counter (or up-down counter
for bipolar coding).

Due to its simple hardware implementation compared to the common arithmetic
operations, SC is very low-cost and energy efficient. But the
traditional SC, however, suffers from long latency and inherent random
fluctuation errors, which are mitigated by the recently proposed
binary interfaced stochastic computing method mentioned below.

\subsection{Counter-based SC multiplication}

\begin{figure}[htp]
\centering
\includegraphics[width=0.9\columnwidth]{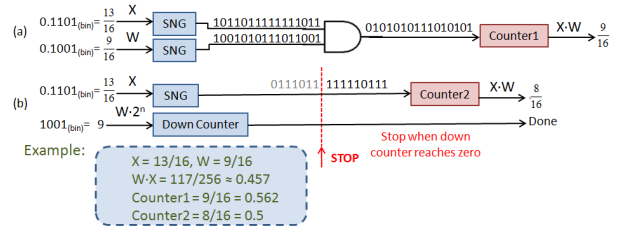}
\caption{(a) Conventional SC multiplier. (b) Counter-based SC multiplier concept\cite{SimLee:DAC'17}.}
\label{fig:cbsc_mul_concept}
\vspace{-0.1in}
\end{figure}

Assume the bit width is $n$ for the given two binary numbers $x$
and $w$. The conventional SC multiplier using {\it AND} gate (for unipolar
encoding) will take $2^n$, which is the length of bit-stream of
stochastic number (SN), cycles to finish the work.  To improve this,
Sim {\it et al.} in~\cite{SimLee:DAC'17} proposed a counter-based SC
multiplier design shown in Fig.~\ref{fig:cbsc_mul_concept}. 
The multiplier mainly
consists of two counters. The down counter counts the binary value of
the input $w$ and the up counter counts the result $x \cdot w$. So the
operation only takes $w \cdot 2^n$ cycles to finish.  One example is given in
Fig.~\ref{fig:cbsc_mul_concept}. More importantly, the
stochastic number of input $x$ can be generated in a deterministic way
without hurting the accuracy (actually more accurate). As a
result, such design is more simple as we eliminate the two
traditional stochastic number generators or SNGs (typically using
Linear Feedback Shift Registers) and {\it AND} gates in exchange of a
down-counter, which is much cheaper than SNG.
%%%%%%%%%%%%%%%%%%%%%%%%%%%%%%%%%%%%%%%%%%%%% 

\begin{figure}[htp]
%\begin{wrapfigure}{L}{0.4\textwidth}
\centering
\includegraphics[width=0.85\columnwidth]{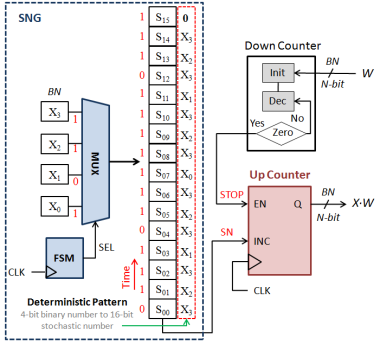}
\caption{Counter-Based SC Multiplier.}
\label{fig:reconf_arsc_multiplier}
%\vspace{-0.1in}
% \end{wrapfigure}
\end{figure}

To generate the stochastic number of $x$, which can be a conventional 
low-discrepancy random number, the authors proposed a deterministic way 
to do this. 
The method evenly distributes the $x_{i-1}$, which is the $i-th$
bit of $x$, based on its binary weight $2^i$. For instance, if $i=3$,
then $x_2$ will appear 4 times in the resulting stochastic number as shown in
Fig.~\ref{fig:cbsc_mul_concept}.  
Such stochastic number generation can be simplified
and implemented by a FSM and a MUX. 
The whole counter based SN multiplication design is shown in
Fig.~\ref{fig:reconf_arsc_multiplier}.

% \begin{figure}
% % \begin{wrapfigure}{L}{0.4\textwidth}
% \centering
% \includegraphics[width=0.38\columnwidth]{./SC_mul_SNG.eps}
% \caption{Existing bit-interfaced SC multiplier and SNG\cite{xxx}.}
% \label{fig:sc_mul_sng}
% %\vspace{-0.1in}
% %\end{wrapfigure}
% \end{figure}

% \begin{figure}
% %\begin{wrapfigure}{L}{0.4\textwidth}
% \centering
% \includegraphics[width=0.38\columnwidth]{./sc_mac.eps}
% \caption{The MAC unit using signed BISC multiplication~\cite{xxx}.}
% \label{fig:bisc_mac}
% % \vspace{-0.1in}
% \end{figure}
% %\end{wrapfigure}

% \begin{figure}[htp]
% \centering
% \includegraphics[width=0.4\columnwidth]{./reconfig_sc_diagram.eps}
% \caption{The proposed reconfigurable SC MAC unit using signed BISC multiplication~\cite{xxx}.}
% \label{fig:reconf_nisc_multiplier}
% %\vspace{-0.1in}
% \end{figure}

% \begin{figure}[htp]
% \centering
% \includegraphics[width=0.4\columnwidth]{./reconf_sc_arch.eps}
% \caption{The proposed reconfigurable SC MAC architecture~\cite{xxx}.}
% \label{fig:reconf_nisc_multiplier}
% %\vspace{-0.1in}
% \end{figure}

%% file: architecture.tex
\section{Proposed run-time ARSC for 2D DCT/IDCT}
\label{sec:architecture}

In this section, we present the proposed accuracy-reconfigurable
stochastic computing (ARSC) method based on the counter-based SC
framework. The key idea is to {\it dynamically adjust the bit-width} of
the coming data for multiplication intensive computing so that we can
reduce the accuracy of the computing progressively using SC. At the
same time, we also reduce effective latency of the computing logic so
that we can compensate aging-induced delay increases. Since we reduce
effective latency of computing logic, we can reduce the frequency
while still being able to maintain the same throughput of the whole
computing process as required by the application.

We will illustrate the proposed ARSC method using an image compression
application based on computing intensive 2D discrete cosine
transformation, DCT and inverse DCT algorithms.

We first briefly review DCT and IDCT computing processes. 
2D discrete cosine transformation (DCT) filter is an effective method for 
eliminating high-frequency noise, by transforming the image data into
spatial frequency domain, masking the high-frequency components, and then
transforming back to the original space domain. A 2D DCT consists of two
separate 1D DCT operations, which can be denoted as
\begin{equation}
    f_k = a_0\sqrt{N} + \sqrt\frac2N \sum^{N-1}_{i=1}a_i \cos\frac{(2i+1)k\pi}{2N},\,0 \le k < N,
    \label{eqn:dct}
\end{equation}
where vector $\{a_i\}$ is the original data, and $\{f_k\}$ is the result of 1D
DCT. A 2D DCT is completed by applying 1D DCT on each column and then on each
row of the matrix. With the image data $T(x,y)$ transformed into its 2D 
frequency domain $F(x,y)$, a filtered frequency map $\mathcal{F}(x,y)$ 
can be obtained by applying a mask
\begin{equation}
    \mathcal{F}(x,y) = F(x,y)m(x,y),
\end{equation}
where $m(x,y)$ is the mask map valued 0 at high frequencies and 1 at low
frequencies. The filtered data $\mathcal{T}(x,y)$ is then obtained by taking 
the inverse 2D DCT on the filtered frequency map $\mathcal{F}(x,y)$. 
Similar to its forward counterpart, the inverse 2D DCT consists of two 
separate inverse 1D DCT steps on the rows and columns respectively. 
The inverse 1D transformation of \eqref{eqn:dct} is
\begin{equation}
    a_i = \frac{f_0}{\sqrt{N}} + \sqrt\frac2N \sum_{k=1}^{N-1}f_k \cos\frac{(2i+1)k\pi}{2N},0 \le i < N.
  \end{equation}
As we can observe, the primary computing in the 2D DCT/IDCT algorithms
are essentially multiply-accumulate operation (MAC). 

\subsection{ARSC architecture for the 2D DCT/IDCT}
\label{sec:re-conf sc arch}

\begin{figure}[!htb]
\centering
\includegraphics[width=0.9\columnwidth]{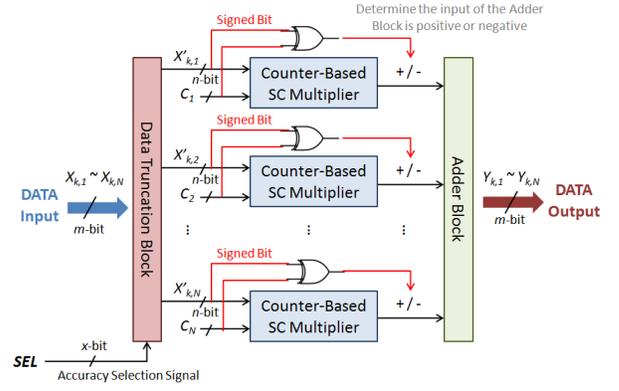}
\caption{The proposed ARSC-based MAC unit}
\label{fig:ARSC_MAC}
\end{figure}

Fig.~\ref{fig:ARSC_MAC} show the proposed ARSC-based MAC unit used in
the DCT/IDCT applications. 

\begin{figure}[!htb]
\centering
\includegraphics[width=0.85\columnwidth]{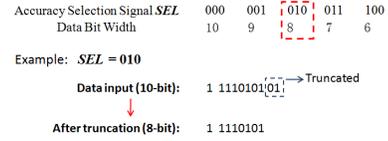}
\caption{Data truncation example.}
\label{fig:data_trun_exp}
\end{figure}

The ARSC MAC unit includes the input data truncation/reconfiguration
block, the multipliers and the adder block. We use the counter-based
SC multiplier to realize SC multiplication as shown in
Fig.~\ref{fig:reconf_arsc_multiplier}. The proposed ARSC module does
the dynamic accuracy arrangement by adjusting the bit width of the
input data that participate in the counter-based SC multiplier in the
ARSC MAC unit, which is realized by the data truncation block. For
instance, when the initial data is $m$-bit, represented in
signed-and-magnitude form, $X_k,_i$ (i = 1, 2, ..., N) in
Fig.~\ref{fig:ARSC_MAC}.  In this line, an x-bit accuracy selection
signal $SEL$ is used to tell the truncation block how many bits it
needs to keep. In our design, for instance, we have 5 states
representing from 10-bit to 6-bit configurations, so x=3 will be enough to
distinguish the 5 states. After data truncation, $X_k,_i$ is
transformed to $X'_k,_i$, which is an truncated $n$-bit binary
number. Fig.~\ref{fig:data_trun_exp} illustrate how the data
truncation block works. Notice that we keep the sign bit, and truncate
the least significant bits from the right side, which is compatible
with the bit-width based progressive SC computing scheme.

After the ARCS MAC process finishes, we will add 0 at the end of the
output number to make it the same bit width as the input binary number
to keep bit-width compatibility between different computing modules.
As SC computing time is directly proportional to the bit-width, or more
precisely proportional to $O(2^{bitwdith})$, one bit-width reduction can
dramatically reduce the SC computing time by half, which can be very
effective for mitigate the aging effects. 

Notice that our counter-based SC multiplier only deals with uni-polar
number multiplication, whose range in $[0,1]$. We keep the sign bits
for all the DCT coefficients $C_i$ (i = 1, 2, ..., N). The sign bit of
$C_i$ and the sign bit of $X'_k,_i$ will perform an $XOR$ operation to
determine whether the product obtained from the counter-based SC
multiplier is positive or negative before participating in the add
operation which is carried out at the adder block.

\begin{figure}[!htb]
\centering
\includegraphics[width=0.85\columnwidth]{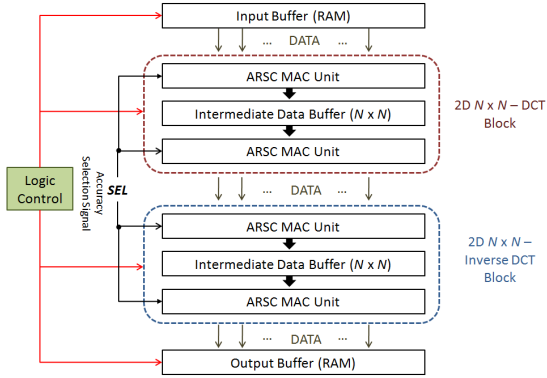}
\caption{The top-level diagram for the ARSC-based image 
compressing/decompressing application.}
\label{fig:2D_DCT_TOP}
\end{figure}

Fig.~\ref{fig:2D_DCT_TOP} shows the architecture of the ARSC consists
of several parts, including the input buffer, two 2D $N$-DCT blocks
($N$ is 8 here), logic control unit, an accuracy selection signal and
an output buffer.  Since SC still has longer computing latency
compared to the conventional arithmetic units, we propose to use the
parallel computing structure to accelerate the process. The input
buffer will send 8 image pixel data to the 2D DCT block once due to
the 8-DCT method. The frameworks of the DCT and the inverse DCT block
actually are the same. The 2D DCT block is made up with two ARSC MAC
units for the 1D DCT process, and an intermediate
buffer. The intermediate buffer is used to save the output data from
the 1D DCT process as the first ARSC MAC unit get the data from input
in row and the second one in column, respectively. The second ARSC MAC
unit will not start working until the intermediate data buffer is
full. And the Logic Control Unit will send the control signal to all
of the blocks to control the data flow of the whole DCT/IDCT process.

%We want the re-configurable SC to process  
%data with different bit width. Firstly, the data need to be truncated to the 
%bit length that we want (ex. from $m$-bit to $n$-bit) by the data truncation 
%module. The "S" is the control signal, telling the module how many bits it need% 
%to keep. For example, if $X$ is the $m$-bit input and $X'$ is the $n$-bit 
%output. The $X$ stands for the image data, which is in the range [0,255]. And 
%$C$ stands for the DCT coefficients, which is in the range [0,1).

%The top-level block diagram of the re-configurable stochastic computing 
%architecture used for 2D DCT (discrete cosine transform) is depicted in 
%Fig.~\ref{fig:2D_DCT_TOP}. The logic control part is used to manage the time 
%sequence of the entire system. The re-configurable architecture is composed of 
%an input buffer (RAM), data truncation module ("TRUN" module in the figure), 2 
%$N$-DCT module, an $N$$\times$$N$ intermediate data buffer, and finally an 
%output buffer for storing the 2D DCT results.

%\textbf{Data Truncation Module.}  

%\textbf{2D $N$$\times$$N$-DCT Module.} Authors in~\cite{} have already proved 
%that 2D DCT is equaled to two 1D DCT process. We use two 1D $N$-DCT module in 
%our proposed architecture, which is highlighted with green dot line in 
%Fig.~\ref{fig:2D_DCT_TOP}, in order to finish the 2D $N$$\times$$N$-DCT %process.
%One, whose name is followed by "(H)", does the 1D DCT in the direction of 
%horizontal. The other, whose name is followed by "(V)", does the DCT process in% 
%the direction of vertical. 

%Obviously, the second $N$-DCT module, which gets the input from the vertical 
%direction, needs to wait until all $N$ 1D DCT process in the horizontal 
%direction to be finished. So we use an intermediate data buffer to store the 
%DCT process results of the first module. By the way, the coefficients used in 
%both DCT processes are the same.

%% file: hard.tex
\section{Hardware Implementation}
\label{sec:hardware}

To evaluate the hardware cost of the re-configurable stochastic computing 
module, including the area, delay and power consumption of the module, 
the proposed design was implemented in Verilog and synthesized 
using Xilinx ISE 14.7 for XC6SLX45 device of Spartan-6 family. 
Different from the ASIC-based module, FPGA mainly 
use the LUT-based operations~\cite{Guo:ASPDAC'20}. So for the design area 
measurement, we simply count the number of LUTs after the module 
is synthesized. As mentioned in Sec.~\ref{sec:architecture}, 
the design totally utilizes 17569 LUTs to support the parallel SC blocks. 
We show the details of the FPGA hardware resource utilization information 
in table~\ref{tab:hard_resource}. To evaluate the power consumption, 
we use the Xilinx Power Estimator downloaded from 
the official website, which can easily obtain the total power consumption.
We'll discuss the power consumption of the design later in 
Sec.~\ref{sec:results}. 
For the delay measurement, 
we obtain the critical path of the ARSC design from the Xilinx ISE 14.7 
timing summary after the design is synthesized, 
showing that the hardware delay, which is calculated from the worst case 
critical path, is 11.348ns. 
It means that the highest frequency the 
ARSC design can run is 88.1M. Since we use the digital clock manager (DCM)
IP of Xilinx ISE 14.7 to generate the clock signal and the input system 
clock of the DCM is 100M for the Spartan-6 family boards. 
The highest frequency DCM can output is 85.7M. 
So we choose this frequency to be the 
initial global clock signal of the ARSC design.

%For the delay measurement, since the Xilinx ISE 14.7 software will give the 
%timing summary after the module being synthesized, we can easily get the data. 
%And to evaluate the power, we use the Xilinx Power Estimator downloaded from 
%the official website, which can easily calculate the total power cost after 
%the required parameter being input.  

%From table.~\ref{tab:hard_cost}, we notice that the delay of our design is 
%11.348ns, which means the highest frequency our design can work on is 
%88.1MHz. We use Xilinx Digital Clock Manager (DCM) to generate the clock 
%frequency we want from the input global 100MHz clock, which is 81.8MHz here. 
%As mentioned in Sec.~\ref{sec:architecture}, %%check
%to accelerate the process, our re-configurable SC DCT-IDCT architecture is 
%designed to work parallelly. The design totally utilizes 17569 LUTs and 8 block 
%RAMs to support 8 parallel SC components.  

\begin{table}[]
\centering
\resizebox{\columnwidth}{!}{%
\begin{tabular}{ccccc}
\hline
\multicolumn{5}{|c|}{\begin{tabular}[c]{@{}c@{}}ARSC Design Hardware Resource Utilization\end{tabular}} \\ \hline
\multicolumn{1}{|c|}{Slice Registers} &
  \multicolumn{1}{c|}{Slice LUTs} &
  \multicolumn{1}{c|}{LUT FF Pairs} &
  \multicolumn{1}{c|}{RAMB16BWERS} &
  \multicolumn{1}{c|}{RAMB8BWERS} \\ \hline
\multicolumn{1}{|c|}{11041} &
  \multicolumn{1}{c|}{17569} &
  \multicolumn{1}{c|}{18098} &
  \multicolumn{1}{c|}{66} &
  \multicolumn{1}{c|}{1} \\ \hline
\end{tabular}%
}
\caption{ARSC design hardware resource utilization.}
\label{tab:hard_resource}
\end{table}

%% file: results.tex
\section{Experimental results and discussion}
\label{sec:results}

In this section, we present results from the proposed ARSC computing
method for the aging mitigation. The proposed ARSC DCT/IDCT image
compression algorithms were implemented on the  Xilinx XC6SLX45 FPGA
platform.

\begin{figure}[!htb]
\centering
\subfigure[]{
\label{fig:subfig:image_a}
\includegraphics[width=0.3\columnwidth]{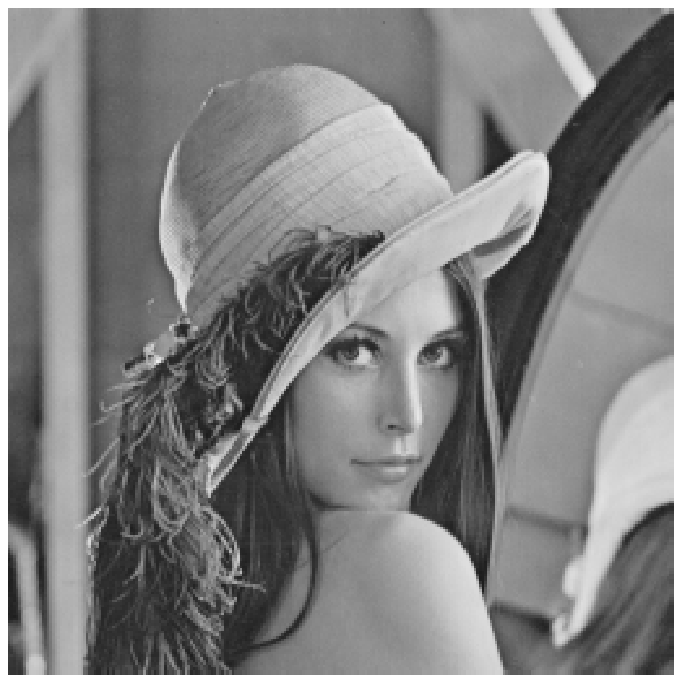}}
\subfigure[]{
\label{fig:subfig:image_b}
\includegraphics[width=0.3\columnwidth]{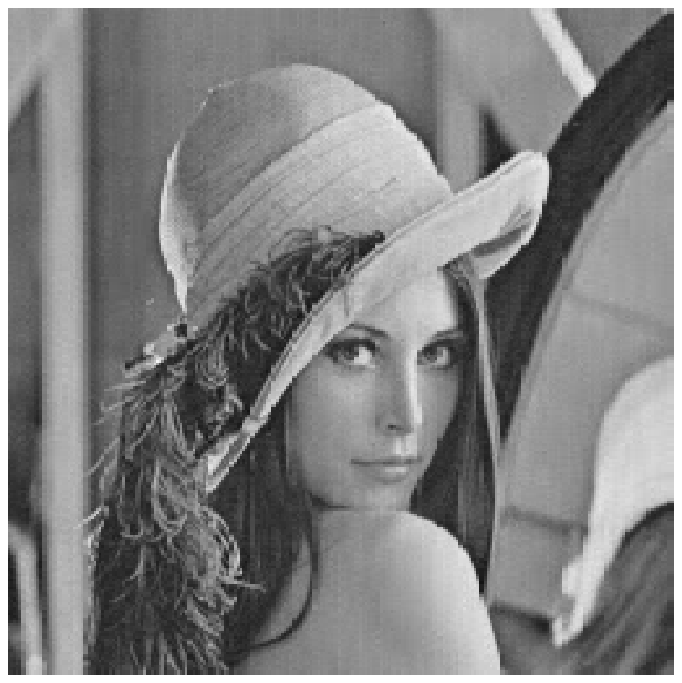}}
\subfigure[]{
\label{fig:subfig:image_c}
\includegraphics[width=0.3\columnwidth]{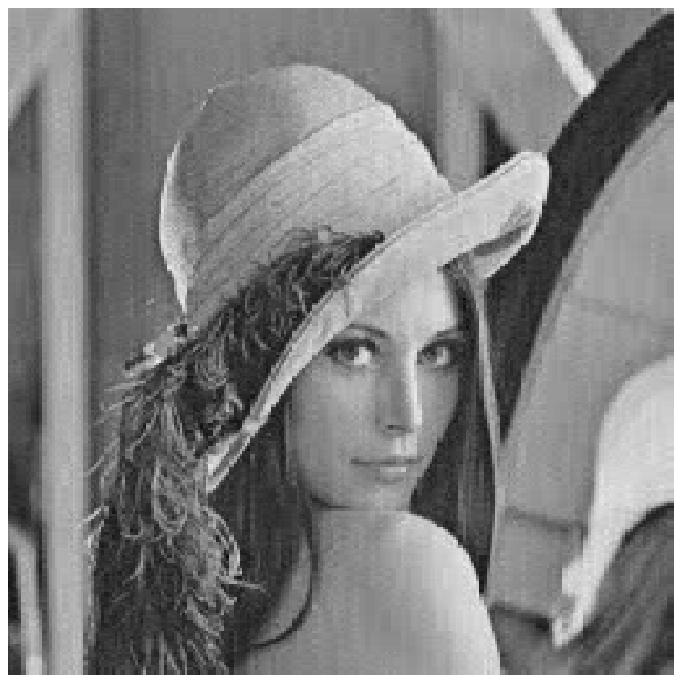}}
\subfigure[]{
\label{fig:subfig:image_d}
\includegraphics[width=0.3\columnwidth]{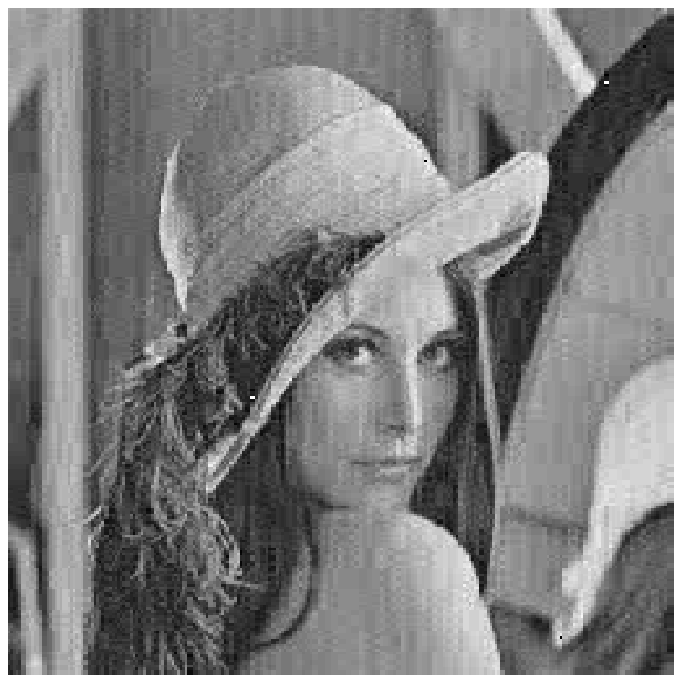}}
\subfigure[]{
\label{fig:subfig:image_e}
\includegraphics[width=0.3\columnwidth]{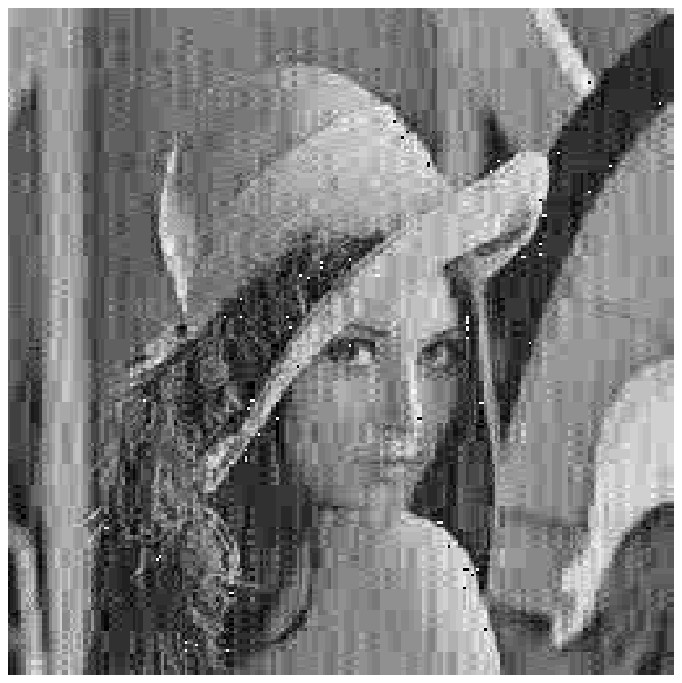}}
\subfigure[]{
\label{fig:subfig:image_f}
\includegraphics[width=0.3\columnwidth]{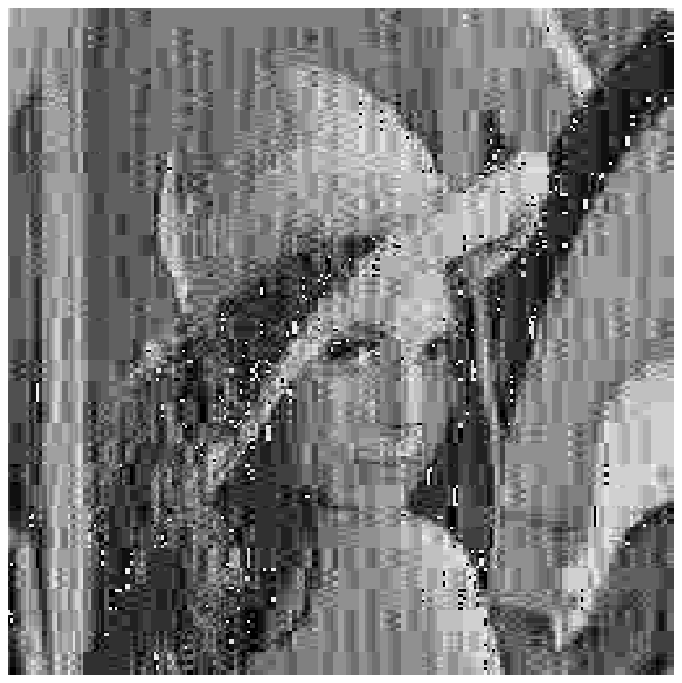}}
\caption{(a) The initial image before DCT/IDCT process;
(b) Image after DCT/IDCT process using 10-bit stochastic multiplication;
(c) 9-bit. (d) 8-bit. (e) 7-bit; (f) 6-bit}
\label{fig:Image_comp_results}
\end{figure}

We first show the image compression results with different accuracy in
Fig.~\ref{fig:Image_comp_results}.
The Fig.~\ref{fig:subfig:image_a} shows the original figure without any
compression. And Fig.~\ref{fig:subfig:image_b} ~
Fig.~\ref{fig:subfig:image_f} show the image quality after the
DCT/IDCT sequence computing with different accuracy, from 10-bit to
6-bit. We use PSNR (Peak Signal to Noise Ratio) to evaluate the accuracy of 
the image compressing/decompressing process, which is shown in 
Table.~\ref{tab:perf_comp}.

\begin{figure}[!htb]
\centering
\includegraphics[width=0.85\columnwidth]{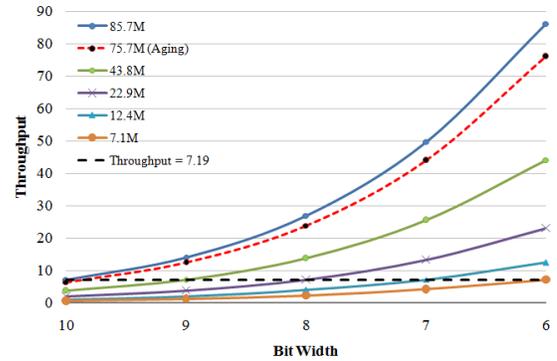}
\caption{Accuracy versus throughput considering the aging.}
\label{fig:bit_vs_thp}
\end{figure}

\begin{figure}[!htb]
\centering
\includegraphics[width=0.85\columnwidth]{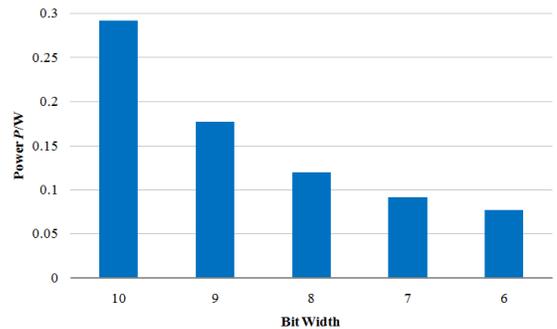}
\caption{Dynamic power arrangement by changing the frequency.}
\label{fig:power_vs_freq}
\end{figure}

If we consider the NBTI induced aging process, which causes the chip
frequency going down over time due to the increased threshed voltage.  
To measure the amount of frequency decrease, we use the time dependent
degradation aware Nangate 45nm standard cell library from Karlsruhe
Institute of Technology (KIT)~\cite{KIT_LIB} to calculate the chip
frequency after 10 years. We use Synopsys design suite to synthesize
the ARSC-based DCT/IDCT design. The timing analysis shows that
frequency of the ASIC-based design of DCT/IDCT will decrease from
1205M to 1064M, as shown in Fig.~\ref{fig:aging}. To simulate the
aging process by FPGA, we simply adjust the frequency output from the
DCM by the same ratio, which is 85.7M to 75.7M (we note that such
mapping may not be perfect as the design technologies used in our ASIC
and FPGA are different). For the DCT/IDCT process, the aging effect
directly affect the throughput. The time of the whole process,
including both of the DCT and IDCT process, is used to calculate the
throughput.

%The throughput, here, means that the number of figures our design can deal with per second. And the whole process 

%Thus, causing the DCT-iDCT process of every image last longer time, and decrease the throughput. 
We show that the throughput with different precision at different
clock frequencies in Fig.~\ref{fig:bit_vs_thp}. The x-axis is the bit
width we use in the stochastic multiplication when we perform the
DCT/IDCT computing.  The y-axis is the throughput, meaning the number
of images the design can deal with per second.
Fig.~\ref{fig:bit_vs_thp} also shows the throughput of 7.19 images per
second for different frequencies and precision with the dashed black
line.  As we can see, initially (when the aging process hasn't start
yet), if we use the full 10-bit precision, the throughput at 85.7MHz clock
frequency is 7.19. When the clock frequency decreases to 43.8MHz, we
can still keep the same throughput if we truncate the precision by
only one bit (from 10 to 9).  Due to the aging process, the throughput
will decrease to 6.35. If we degrade the precision by one bit (9-bit),
the throughput will increase to 12.42, which obviously, is larger than
the initial throughput. The red line of dashes in
Fig.~\ref{fig:bit_vs_thp} shows this very clearly. And, by decreasing
the precision of SC multiplication to 6-bit, the throughput will be
about 12 times of the 10-bit precision, which shows huge space we can
mitigate the aging effects if such accuracy is still accepted in
practical applications.

Due to the difficulty of obtaining the hardware delay of the scenarios
in which the data bit width is not 10-bit during SC computing process,
we use the effective latency to evaluate the timing performance of our
design.  The effective latency of the ARSC design is actually the
inverse of the throughput, since it represents the time interval
between the input and the output. We show the latency at the 5th
column of Table~\ref{tab:perf_comp}.

From Table~\ref{tab:perf_comp}, we show that our design
can do the dynamic power management by adjusting the working frequency
as well.  By doing the trade off between the throughput and the
accuracy mentioned before, we can keep the throughput by sacrificing
accuracy when the frequency is cut down due to some low power
consumption requirement situation. For example, if we want to keep the
throughput as 7.19 here, when the power goes down, the frequency of
the ARSC design will also decrease. Thus, the precision (bit width) of
the data our proposed ARSC design can work will also decrease. We show
this clearly in Fig.~\ref{fig:power_vs_freq}.  We notice that we can
save near 74\% of power consumption by sacrificing 10.67dB of the
accuracy loss. We also observed that the proposed ARSC computing framework 
allows much aggressive frequency scaling, which can lead to order of
magnitude power savings compared to the traditional dynamic voltage and
frequency scaling (DVFS) techniques.

\begin{table}[]
\centering
\resizebox{\columnwidth}{!}{%
\begin{tabular}{|c|c|c|c|c|}
\hline
\begin{tabular}[c]{@{}c@{}}Bit Width\end{tabular} & Frequency (MHz) & Power (W) & PSNR (dB) & latency (s) \\ \hline
10 & 85.7 & 0.292 & 38.12 & 0.139 \\ \hline
9  & 43.8 & 0.177 & 34.68 & 0.071 \\ \hline
8  & 22.9 & 0.120 & 31.27 & 0.037 \\ \hline
7  & 12.4 & 0.092 & 28.70 & 0.020 \\ \hline
6  & 7.1  & 0.077 & 27.45 & 0.012 \\ \hline
\end{tabular}%
}
%\vspace{-0.2cm}
\caption{Key performance metric comparison under the same throughput.}
\label{tab:perf_comp}
\end{table}